\documentclass[twocolumn]{aastex631}

\usepackage{CJK}

\begin{document}

\title{Ejecta Evolution Following a Planned Impact into an Asteroid: \\The First Five Weeks}

\shorttitle{DART Ejecta Evolution}
\shortauthors{Kareta et al.}

\author[0000-0003-1008-7499]{Theodore Kareta}
\affiliation{Lowell Observatory\\
Flagstaff, AZ, USA}

\author[0000-0003-3091-5757]{Cristina Thomas}
\affiliation{Northern Arizona University}

\begin{CJK*}{UTF8}{gbsn}
\author[0000-0003-3841-9977]{Jian-Yang Li (李荐扬)}
\affiliation{Planetary Science Institute}
\author[0000-0003-2781-6897]{Matthew M. Knight}
\affiliation{Physics Department\\
United States Naval Academy\\
572C Holloway Rd, Annapolis, MD 21402, USA}

\author{Nicholas Moskovitz}
\affiliation{Lowell Observatory\\
Flagstaff, AZ, USA}

\author[0000-0003-2341-2238]{Agata Ro{\.z}ek}
\affiliation{Institute for Astronomy, University of Edinburgh, Royal Observatory, Edinburgh, EH9 3HJ, UK}

\author{Michele T. Bannister}
\affiliation{University of Canterbury}

\author{Simone Ieva}
\affiliation{INAF - Osservatorio Astronomico di Roma}

\author{Colin Snodgrass}
\affiliation{Institute for Astronomy, University of Edinburgh, Royal Observatory, Edinburgh, EH9 3HJ, UK}

\author{Petr Pravec}
\affiliation{Astronomical Institute, Academy of Sciences of the Czech Republic}

\author{Eileen V. Ryan}
\affiliation{New Mexico Institute of Mining and Technology/Magdalena Ridge Observatory\\
801 Leroy Place, Socorro, New Mexico, USA, 87801}

\author{William H. Ryan}
\affiliation{New Mexico Institute of Mining and Technology/Magdalena Ridge Observatory\\
801 Leroy Place, Socorro, New Mexico, USA, 87801}

\author[0000-0003-1391-5851]{Eugene G. Fahnestock}
\affiliation{Jet Propulsion Laboratory\\
California Institute of Technology}

\author{Andrew S. Rivkin}
\affiliation{Johns Hopkins University Applied Physics Laboratory}

\author{Nancy Chabot}
\affiliation{Johns Hopkins University Applied Physics Laboratory}

\author{Alan Fitzsimmons}
\affiliation{Queen's University Belfast}

\author{David Osip}
\affiliation{Las Campanas Observatory}

\author{Tim Lister}
\affiliation{Las Cumbres Observatory}

\author{Gal Sarid}
\affiliation{SETI Institute}

\author{Masatoshi Hirabayashi}
\affiliation{Daniel Guggenheim School of Aerospace Engineering\\
Georgia Institute of Technology\\
620 Cherry Street, Atlanta, Georgia, USA, 30332}

\author{Tony Farnham}
\affiliation{University of Maryland}

\author{Gonzalo Tancredi}
\affiliation{Depto. Astronomía, Udelar, Uruguay}

\author{Patrick Michel}
\affiliation{Université Côte d’Azur, Observatoire de la Côte d’Azur, CNRS, Laboratoire Lagrange, Nice, France}
\affiliation{The University of Tokyo, Department of Systems Innovation, School of Engineering, Tokyo, Japan}
 
\author{Richard Wainscoat}
\affiliation{University of Hawai'i}

\author[0000-0002-0439-9341]{Rob Weryk}
\affiliation{Physics and Astronomy\\
University of Western Ontario\\
1151 Richmond Street, London ON, N6A 3K7, Canada}

\author{Bonnie Burrati}
\affiliation{Jet Propulsion Laboratory\\ California Institute of Technology}

\author{Jana Pittichová}
\affiliation{Jet Propulsion Laboratory\\ California Institute of Technology}

\author{Ryan Ridden-Harper}
\affiliation{University of Canterbury}

\author{Nicole J. Tan}
\affiliation{University of Canterbury}

\author{Paul Tristram}
\affiliation{MOA}

\author{Tyler Brown}
\affiliation{University of Canterbury}

\author{Mariangela Bonavita}
\affiliation{Institute for Astronomy, University of Edinburgh, Royal Observatory, Edinburgh, EH9 3HJ, UK}

\author{Martin Burgdorf}
\affiliation{Universität Hamburg}

\author{Elahe Khalouei}
\affiliation{Seoul National University}

\author{Penelope Longa}
\affiliation{Universidad de Antofagasta}

\author{Markus Rabus}
\affiliation{Universidad Católica de la Santísima Concepción}

\author{Sedighe Sajadian}
\affiliation{Isfahan University of Technology}

\author{Uffe Graae Jorgensen}
\affiliation{University of Copenhagen}

\author{Martin Dominik}
\affiliation{University of St. Andrews}

\author{Jean-Baptiste Kikwaya}
\affiliation{Vatican Observatory}

\author{Elena Mazzotta Epifani}
\affiliation{INAF - Osservatorio Astronomico di Roma}

\author{Elisabetta Dotto}
\affiliation{INAF - Osservatorio Astronomico di Roma}

\author{J.D. Prasanna Deshapriya}
\affiliation{INAF - Osservatorio Astronomico di Roma}

\author{Pedro H. Hasselmann}
\affiliation{INAF - Osservatorio Astronomico di Roma}

\author{Massimo Dall’Ora}
\affiliation{INAF - Osservatorio Astronomico di Capodimonte}

\author{Lyu Abe}
\affiliation{Université Côte d’Azur, Observatoire de la Côte d’Azur, CNRS, Laboratoire Lagrange, Nice, France}
\author{Tristan Guillot}
\affiliation{Université Côte d’Azur, Observatoire de la Côte d’Azur, CNRS, Laboratoire Lagrange, Nice, France}
\author{Djamel Mékarnia}
\affiliation{Université Côte d’Azur, Observatoire de la Côte d’Azur, CNRS, Laboratoire Lagrange, Nice, France}
\author{Abdelkrim Agabi}
\affiliation{Université Côte d’Azur, Observatoire de la Côte d’Azur, CNRS, Laboratoire Lagrange, Nice, France}
\author{Philippe Bendjoya}
\affiliation{Université Côte d’Azur, Observatoire de la Côte d’Azur, CNRS, Laboratoire Lagrange, Nice, France}
\author{Olga Suarez}
\affiliation{Université Côte d’Azur, Observatoire de la Côte d’Azur, CNRS, Laboratoire Lagrange, Nice, France}
\author{Amaury Triaud}
\affiliation{University of Birmingham}

\author{Thomas Gasparetto}
\affiliation{INAF - Trieste}

\author{Maximillian N. Günther}
\affiliation{European Space Agency}

\author{Michael Kueppers}
\affiliation{European Space Agency}

\author{Bruno Merin }
\affiliation{European Space Agency}

\author{Joseph Chatelain}
\affiliation{Las Cumbres Observatory}

\author{Edward Gomez}
\affiliation{Las Cumbres Observatory}

\author{Helen Usher}
\affiliation{Open University}

\author{Cai Stoddard-Jones}
\affiliation{Cardiff University}

\author{Matthew Bartnik}
\affiliation{Michigan State University}

\author{Michael Bellaver}
\affiliation{Michigan State University}

\author{Brenna Chetan}
\affiliation{Michigan State University}

\author{Emma Dugan}
\affiliation{Michigan State University}

\author{Tori Fallon}
\affiliation{Michigan State University}

\author{Jeremy Fedewa}
\affiliation{Michigan State University}

\author{Caitlyn Gerhard}
\affiliation{Michigan State University}

\author{Seth A. Jacobson}
\affiliation{Michigan State University}

\author{Shane Painter}
\affiliation{Michigan State University}

\author{David-Michael Peterson}
\affiliation{Michigan State University}

\author{Joseph E. Rodriguez}
\affiliation{Michigan State University}

\author{Cody Smith}
\affiliation{Michigan State University}

\author{Kirill V. Sokolovsky}
\affiliation{Michigan State University}

\author{Hannah Sullivan}
\affiliation{Michigan State University}

\author{Kate Townley}
\affiliation{Michigan State University}

\author{Sarah Watson}
\affiliation{Michigan State University}

\author{Levi Webb}
\affiliation{Michigan State University}

\author{Josep M. Trigo-Rodríguez}
\affiliation{Institute of Space Sciences (CSIC-IEEC)\\
08193 Cerdanyola del Vallés (Barcelona)\\
Carrer de Can Magrans, s/n, Catalonia, Spain}

\author{Josep M. Llenas}
\affiliation{Pujalt Obs., Barcelona, Catalonia, Spain}

\author{Ignacio Pérez-García}
\affiliation{Instituto de Astrofísica de Andalucía, IAA-CSIC, Granada, Spain}

\author{A.J. Castro-Tirado}
\affiliation{Instituto de Astrofísica de Andalucía, IAA-CSIC, Granada, Spain}

\author{Jean-Baptiste Vincent}
\affiliation{DLR Institute of Planetary Research, Germany}

\author{Alessandra Migliorini}
\affiliation{INAF - Institute for Space Astrophysics and Planetology}

\author{Monica Lazzarin}
\affiliation{Dipartimento di Fisica e Astronomia, Padova University, Italy}

\author{Fiorangela La Forgia}
\affiliation{Dipartimento di Fisica e Astronomia, Padova University, Italy}

\author{Fabio Ferrari}
\affiliation{Politecnico di Milano, Italy}

\author{Tom Polakis}
\affiliation{Lowell Observatory\\
Flagstaff, AZ, USA}

\author{Brian Skiff}
\affiliation{Lowell Observatory\\
Flagstaff, AZ, USA}

\begin{abstract}

The impact of the DART spacecraft into Dimorphos, moon of the asteroid Didymos, changed Dimorphos’ orbit substantially, largely from the ejection of material. We present results from twelve Earth-based facilities involved in a world-wide campaign to monitor the brightness and morphology of the ejecta in the first 35 days after impact. After an initial brightening of $\sim1.4$ magnitudes, we find consistent dimming rates of $0.11-0.12$ magnitudes/day in the first week, and $0.08-0.09$ magnitudes/day over entire the study period. The system returned to its pre-impact brightness $24.3-25.3$ days after impact though the primary ejecta tail remained. The dimming paused briefly eight days after impact, near in time to the appearance of the second tail. This was likely due to a secondary release of material after re-impact of a boulder released in the initial impact, though movement of the primary ejecta through the aperture likely played a role. 

\end{abstract}


\section{Introduction} \label{sec:intro}
At 23:14 UTC September 26th, 2022, the Double Asteroid Redirection Test (DART) spacecraft impacted Dimorphos, the moon of the asteroid (65803) Didymos \citep{2023Natur.616..443D}. The goal of DART \citep{2021PSJ.....2..173R} was to measurably change the orbital period of the moon and thus derive an estimate of the momentum enhancement factor $\beta$ \citep{2023Natur.616..457C} i.e., how much momentum was imparted to the body compared to how much momentum the spacecraft had prior to impact. Understanding how well a kinetic impactor can alter an asteroid’s orbit is key to knowing how effective that technique would be in future planetary defense situations. The value of $\beta$ was expected to be above one due to the ejection of material from the surface, broadly in the anti-impact direction \citep{2022PSJ.....3..248S}. Measuring and interpreting the properties of the ejecta and how it evolved over time therefore contributes to a broader understanding of not just how well DART worked, but \textit{why}.

Early results (see, e.g., \citealt{2023Natur.616..457C, 2023Natur.616..443D, 2023Natur.616..448T, 2023Natur.616..452L, 2023Natur.616..461G, 2023NatCo..14.3055D}) have shown that DART was highly successful at altering the orbit of Dimorphos around Didymos. The estimated value of $\beta=3.6$, assuming the densities of Dimorphos and Didymos are the same \citep{2023Natur.616..457C}, is significantly higher than one, which is consistent with the rapid appearance of ejecta seen by the Hubble Space Telescope (HST) \citep{2023Natur.616..452L} and ground-based observers (see, e.g., \citealt{2023Natur.616..461G, 2023A&A...671L..11O}), immediately indicating that significant material was ejected and ejected rapidly. \citealt{2023Natur.616..452L} found the structure of the ejecta to be highly complex, driven first by dynamical interactions with Didymos and Dimorphos, and later by radiation pressure. The interaction of the ejecta with the binary system, especially any possible re-accretion of material previously lofted, is a key aspect of measuring and interpreting $\beta$, at least until the ESA Hera mission \citep{2022PSJ.....3..160M} provides a precise measurement of the actual mass of Dimorphos.

This study presents new ground-based telescopic data from September 26 through November 1st. The data presented expands upon previously-reported ejecta results \citep{2023Natur.616..452L, 2023Natur.616..461G, 2023A&A...671L..11O, 2023ApJ...945L..38B} which have shown the evolution of the ejecta to be a significantly more complex and long-lasting process than anticipated prior to impact. This work combines data from twelve Earth-based facilities that were collected with denser temporal sampling and with wider fields of view than the other published studies. This is a subset of all of the lightcurve campaign observations which are detailed in \citet{Nick_subd}. Previous spacecraft missions like Deep Impact \citep{2005Sci...310..258A} and EPOXI were also complimented by large ground-based \citep{2005Sci...310..265M, 2011ApJ...734L...1M} campaigns to great effect. The ground-based campaign was designed to measure a change in Dimorphos’ orbital period about Didymos, and thus constrain the momentum enhancement factor $\beta$ \citep{2023Natur.616..448T}. Further refinements and deeper analyses of these and other datasets are still ongoing. Continued monitoring of the system by globally distributed telescopes has also allowed a new understanding of the complex evolution of the ejecta.
	
Additionally, DART’s impact provides a unique opportunity to study an  “activated” asteroid (see, e.g., \citealt{2023Natur.616..461G}) created with precisely known impact conditions, hence providing for the first time a clear knowledge of the behavior and evolution of other impact-driven active asteroids. Indeed, impacts are one of the most dominant processes in the Solar System \citep{1989icgp.book.....M}, and yet our knowledge of their mechanics on small bodies are limited by the lack of knowledge about the (natural) impactor and target properties and the lack of observations in the very early stage of ejecta and tail evolution due to their serendipitous discovery. For example, in 2010 (596) Scheila underwent non-repeating but significant mass loss (see, e.g., \citealt{2011ApJ...733L...3B}) that was largely interpreted as impact-driven. Moreover, Scheila’s ejecta coma displayed considerable morphological evolution afterwards \citep{2011ApJ...741L..24I}. A clearer understanding of ejecta resulting from impacts onto asteroids will be increasingly useful to interpret and model (see, e.g., \citealt{2021MNRAS.503.1070L}) impact-driven mass loss as epochs of activity are discovered closer to their start in the era of the Vera Rubin Observatory’s Legacy Survey of Space and Time \citep{2019ApJ...873..111I}.

In this work, we present observations from twelve groundbased observatories on three continents and Hawai'i of the evolution of the Didymos-Dimorphos system in the first five weeks after the DART impact. In Section \ref{sec:methods}, we present a brief overview of our data reduction procedures and how they varied from observatory to observatory. In Section \ref{sec:results}, we present our photometric results, with a specific focus on the photometric behavior of the system eight days after impact, within the context of our and others' deep imaging work. In Section \ref{sec:disc}, we propose several hypotheses to explain the evolution of the brightness of the system over time and begin to apply some of what was learned from DART to deepen our knowledge of other active asteroids.

\section{Observatories and Methods} \label{sec:methods}
\subsection{Observatories}
This work presents photometric and imaging observations from twelve separate observatories from Hawai'i to Antarctica. The Lowell Discovery Telescope (LDT) and Southern Observatory for Astrophysical Research (SOAR) contributed both deep imaging and observations of the long-term photometric dimming. The ASTEP telescope in Antarctica, the Henrietta Swope (Las Campanas) telescope, the Las Cumbres Observatory Global Telescope (LCOGT) network, the Ōtehīwai Mt. John observatory, the Danish 1.54-m (La Silla), the Michigan State University (MSU) campus telescope, and the Lowell Observatory 42”-inch Hall Telescope contributed photometric observations. The Very Large Telescope (VLT), Magdalena Ridge Observatory (MRO), and Canada-France-Hawai’i (CFHT) telescope all contributed deep imaging. Observations from the photometry-only telescopes (e.g. the Lowell 42" or the Swope telescope) were primarily obtained as part of the DART lightcurve campaign \citep{2023Natur.616..448T}, and thus have cadences $<180$ s per exposure with series of exposures regularly spaced over many hours. In contrast, imaging observations were focused on ejecta morphology, generally at longer and less regular cadences. Imaging data from the CFHT and SOAR datasets were obtained through separate deep imaging focused proposals to those telescopes, the VLT images were originally acquisition images from a spectroscopy focused proposal, and the LDT datasets came from a combination of specific imaging and specific lightcurve efforts.

This paper focuses primarily on the bulk photometric behavior of the system in the first month after impact. The deeper imaging efforts will be used to help contextualize these behaviors within the context of the morphological evolution of the ejecta and within the context of the other previously mentioned imaging campaigns, including those at significantly higher resolution or those with sparser time coverage. We discuss how each of these overlapping datasets were reduced and utilized in the following subsections.

\subsection{Data Reductions and Analysis Methods}
This work presents and analyzes ground-based telescopic observations made primarily with Charge Coupled Devices (CCDs). We followed standard procedures in the reduction of each dataset (see, e.g., \citealt{2023Natur.616..448T}), specifically the removal of a median bias frame and division by a normalized flatfield. For datasets where dark counts were notable, dark frames of the same exposure lengths were used to remove those counts. Most of the photometric observations and some of the imaging-focused observations were designed primarily to measure the lightcurve of the system, and thus measure the timing of mutual events.

For a typical dataset, the images were registered (e.g., had their pixel locations associated with sky coordinates) by comparison against patterns of field stars whose positions were recorded accurately by all-sky surveys, most commonly GAIA \citep{2018A&A...616A...1G}. Registration and extraction of sources, commonly using SourceExtractor \citep{1996A&AS..117..393B}, were usually completed as one step. Extracted source brightnesses were compared to calibrated source brightnesses from all-sky surveys covering portions of the sky where Didymos was identified during the study period, most commonly SKYMAPPER \citep{2018PASA...35...10W} and PanSTARRS \citep{2012ApJ...750...99T} or APASS \citep{2016yCat.2336....0H}. For many observers, every one of these steps after basic image reduction was handled by an all-in-one pipeline like PhotometryPipeline \citep{2017A&C....18...47M}. As one example, the Michigan State University (MSU) data were reduced and had photometry extracted using a SourceExtractor-based pipeline called VaST \citep{2018A&A...616A...1G}, and their images were registered (often referred to as “plate-solved”) into sky-coordinates using Astrometry.net \citep{2010AJ....139.1782L} before calibration against APASS. The vast majority of our photometry was obtained through the Sloan r or Johnson R filters. The full list of observatories which contributed photometry to this effort, as well as filters they used and the date ranges they observed over, is available in the Appendix.

While an aperture radius corresponding to a physical length might be preferred for some kinds of analysis, the highly variable nature of the seeing and conditions at the many observatories included in this work lead us to obtain photometry with apertures of a fixed angular size. For almost all of our photometric observers, a radius of 5.0 arcseconds was used to measure the brightness of the inner ejecta. At sites where Didymos remained relatively close to the horizon (e.g, Antarctica, Michigan),  larger apertures were used to account for the correspondingly larger image PSFs. Variations in the aperture size can change measured quantities (dimming rates especially) significantly. As a result, the calibrated photometry from the Danish 1.54m and the Swope telescopes were extracted with multiple different aperture sizes (see Figure \ref{fig:2}) to assess the importance of these effects.

We also note that two photometric datasets -- the one from the Danish 1.54-m (the 2.5 arcsecond reduction) and the Lowell Observatory 42” -- had their photometry cleaned of potential mutual event contamination. This manifested in those two datasets being slightly brighter than contemporaneous observations from other observers, but this effect is small compared to the broad-scale photometric evolution under consideration in this work and we do not discuss it further.

To construct the image stacks that were used for the morphological analysis, the best subset (often excluding those images where field stars were too close to the inner ejecta or tail) were stacked using automated routines which queried for the asteroid system’s non-sidereal rates and rotated the images accordingly.

After stacking and alignment, we first inspected the image stacks using a variety of image scalings. This was used to understand the broad-scale evolution of the ejecta (fans, tails, and faster material) for qualitative comparison with the HST \citep{2023Natur.616..452L} and MUSE \citep{2023A&A...671L..11O} datasets, and to select which ones might be apt for studies of the tail. For the first approximately three weeks after impact, the tail was dominated by grains whose behavior was dominated by radiation pressure and thus the tail was exactly in the anti-Sunward direction (see the legend of Figure 3), but in the last $\sim$10 days of our study the tail started to deviate from being purely anti-Sunward. This is indicative of the tail being composed of successively larger grains throughout the study period, as the smaller grains would have been accelerated away by radiation pressure quickly and larger grains would have been slower to leave the system and would take longer to have their motions become strictly anti-Sunward. This is consistent with the color evolution of the coma reported by other observers \citep{2023A&A...671L..11O, 2023Natur.616..461G}.

\section{Results} \label{sec:results}
\subsection{Bulk Dimming and the `Eight Day Pause'}
\begin{figure}[ht!]
\plotone{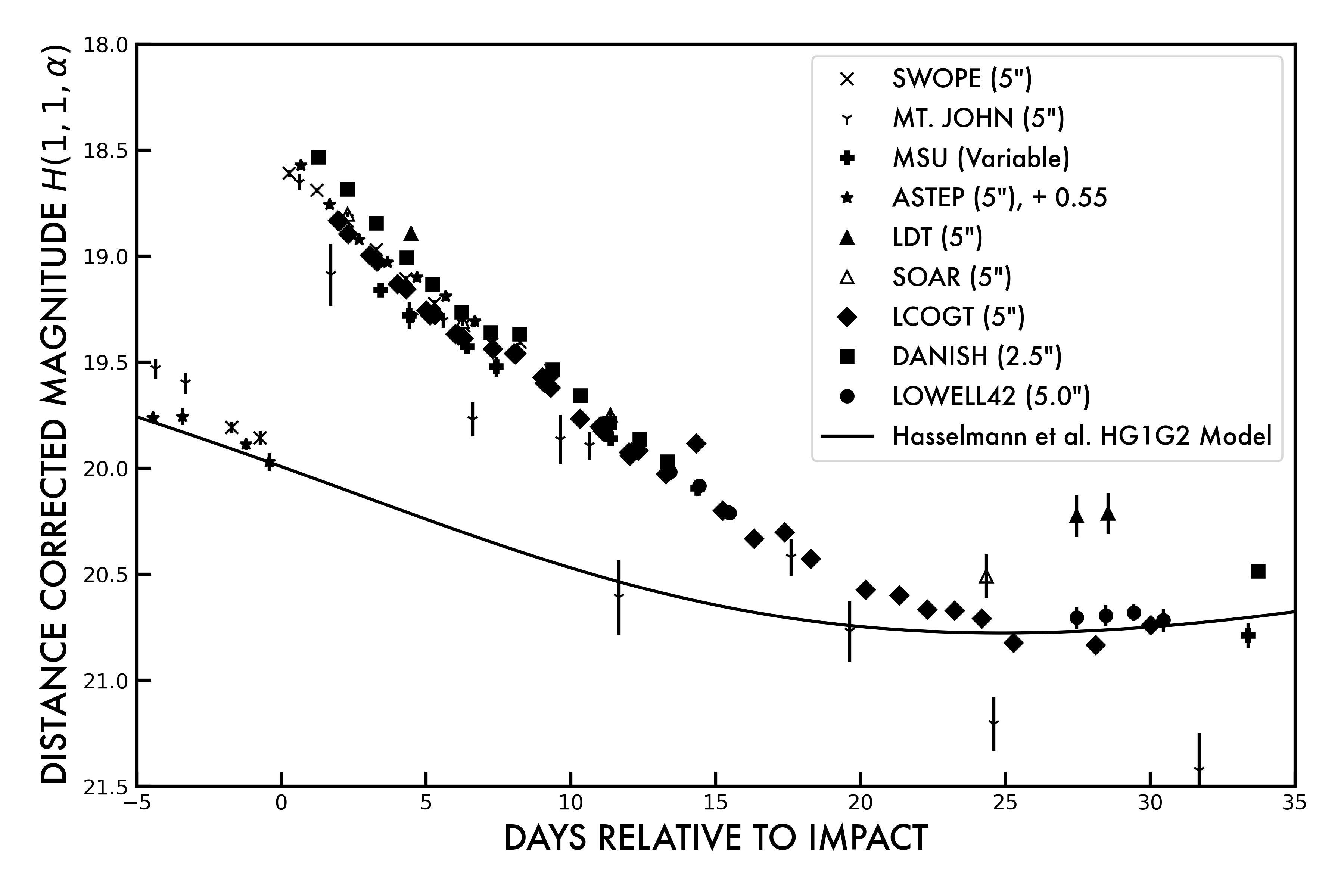}
\caption{The evolution of the brightness of the Didymos system plus the ejecta from the DART impact over time measured from several days before impact to approximately one month after. Different observatories are given their own symbol, and each observation is plotted with errorbars -- though these are frequently smaller than the markers themselves. The dimming was more rapid at first and slowed over time. These brightness measurements were only corrected for heliocentric and geocentric distances, but we also plot an HG1G2 model \citep{2010Icar..209..542M} of the system’s brightness using in-situ derived values (H=18.16, G1=0.84, G2=0.05, \citealt{2023LPICo2806.2023H}) as a black line for context. Thisindicates that some of the dimming is due to phase angle effects, but certainly not all. Only the Danish (squares) and Lowell 42” (circles) have had mutual events removed from their nightly averages, which may move them slightly above contemporaneous observations. A pause in dimming at seven-to-eight days post impact (highlighted in Figure 2) and the time-dependence of the dimming makes ascertaining the bulk wavelength dependence of the dimming rates challenging.}
\label{fig:1}
\end{figure}

First, we present and discuss our observations of the bulk dissipation of the ejecta in the days and weeks after impact and make brief comparisons between them. For the datasets characterized by multiple nights of photometrically calibrated observations, we measured the brightness of the Didymos system and the ejecta with standard aperture photometry, displayed and compared in Figure \ref{fig:1}. For most datasets, we extracted photometry in a 5.0 arcsecond radius, which corresponds to a physical scale of 275 km on September 26th to 400 kilometers on November 1st. Observations from the Danish 1.54-m telescope were extracted in a 2.5 arcsecond radius (scales of 137 to 200 km) and observations from the ASTEP and MSU observatories were extracted with larger variably-sized aperture aiming to capture as much light as possible in variable seeing at those sites. All reported magnitudes are all-night averages to smooth out variations due to the lightcurve of the system. This does not account for all variation (some nights will have `seen' more peaks in the lightcurve than others), but this is a minor effect as can be seen below.

The photometric behavior is qualitatively and quantitatively very similar between different observers. First, a large increase in brightness on the order of $1.4$ magnitudes is evident immediately after impact in all datasets. This brightness increase does not include the $\sim2.3$ magnitude `flash' of brightness seen in the observations by \citet{2023Natur.616..461G}, but instead corresponds to the plateau of brightness enhancement that stabilized within a few hours after impact.

After the initial increase in brightness, the system began to slowly dim over the next few weeks, approaching the pre-impact brightness between $24.3$ and $25.3$ days after impact based on when the LCOGT data cross the predicted brightness curve (an HG1G2 model \citep{2010Icar..209..542M} with parameters derived by \citet{2023LPICo2806.2023H}, shown in Figure \ref{fig:1} as a black line).
LCOGT is the only dataset we have with temporally dense enough coverage to bracket the return to pre-impact brightness, but a linear fit between the two nearest points of the Mt. John dataset produce a similar but significantly less precise estimate. (These data are noisier, so different interpretations of the data are allowed.) To encompass the uncertainty in when the return to expected brightness really happened based on our multiple datasets, we report this as $T+25\pm1$ days post-impact. A linear fit to the magnitudes early on, when dimming was more rapid, would produce a shorter estimate. This is only $\sim1\sigma$ later than the estimate of \citet{2023Natur.616..461G} ($23.7\pm0.7$ days) based on a linear fit to observations early on and several close to when the system returned to its pre-impact brightness, which we view as good agreement despite differences in techniques.

\begin{figure}[ht!]
\plotone{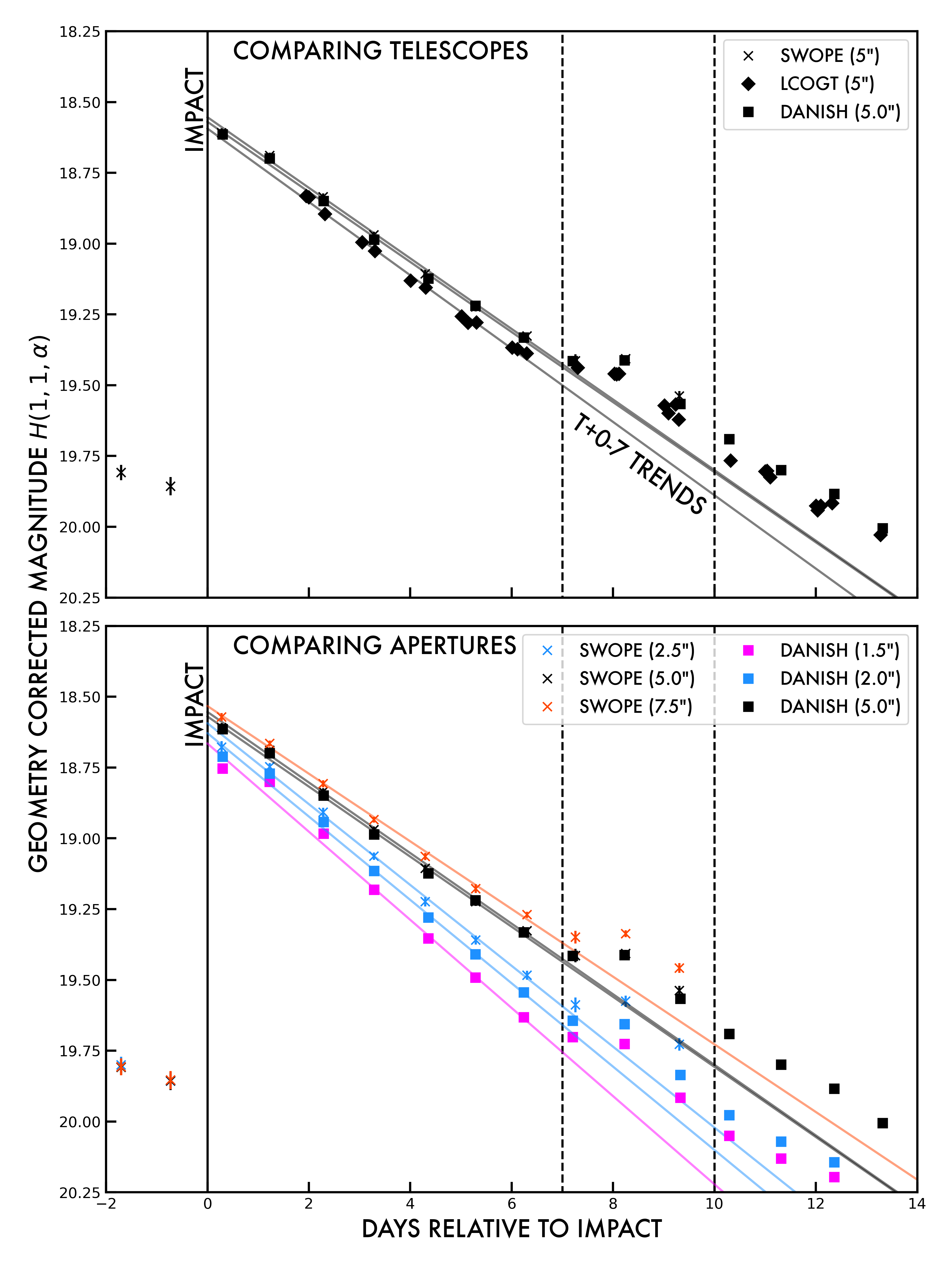}
\caption{Top: A zoomed-in section of Figure 1 highlighting the change in photometric dimming in the Swope (‘x’s), LCOGT (diamonds), and Danish 1.54-m (squares) datasets which have fast enough cadences to detect and characterize it. The dimming appears to “pause” 7-8 days after impact and then continue approximately a day or two later at very similar rates (often within $1-\sigma$ of each other). The Danish dataset had mutual event contamination masked out, while Swope and LCOGT did not. This suggests that mutual events do not play a significant role in this photometric event. The three sites also used slightly different filters which each peak at red wavelengths (Swope utilized Sloan r, LCOGT used the broadband PanSTARRS w (close to Sloan g+r+i), and the Danish 1.54-m utilized the Johnson R filter. While the retrieved dimming rates for identical apertures are really very similar, this may explain some of the vertical offsets between the observatories. Bottom: The same timeframe but focusing on photometry with different aperture radii from the Swope and Danish telescopes, with aperture size indicated by color (pink for 1.5", blue for 2-2.5", black for 5.0", and orange for 7.5"). Smaller apertures dim faster, but the change in photometric behavior at the eight-day mark is similar among aperture sizes.}
\label{fig:2}
\end{figure}

An exception to the monotonic dimming of the ejecta was seen roughly a week after the impact. The ``eight day pause” is clearly identified in the Swope, Danish 1.54-m, and LCOGT datasets and highlighted in Figures \ref{fig:1} and \ref{fig:2} as a temporary pause or slight increase in brightness of the ejecta super-imposed on the overall dimming. The increase is approximately $\sim0.2$ magnitudes at peak and lasts approximately 1-2 days before dimming resumes at similar rates to before the event. The duration and temporal dependence of the photometric event prevents it being due to contamination by a passing field star, and the detection of it by multiple telescopic facilities rules out it being a reduction artifact. This occurs approximately one week after impact at around the same time the ``double tail” was seen in the HST and MUSE datasets \citep{2023Natur.616..452L, 2023A&A...671L..11O}. In the bottom panel of Figure \ref{fig:2}, we explore the aperture size dependence of the dimming in the Swope and Danish 1.54-m datasets. The dimming continues after this photometric event, but the whole system remained brighter than would be expected from the pre-event trends alone. While the smaller apertures dim faster than the larger ones, the difference between successively larger apertures is diminishing.

The slight brightening or pause in dimming appears to have the same ``strength” (i.e., magnitudes brighter than pre-pause trends) between apertures, and at the time resolution allowed by these all-night averages appear to happen at the same time with the same strength across apertures. If material were drifting into the photometric aperture, for example due to projection effects from changing viewing geometry, one might infer earlier brightening in the largest apertures compared to the smaller ones, but this appears to be minimal or not the case. We discuss possible origin scenarios and their relationship with other observables (e.g. color changes, morphological evolution) in the Discussion section.

The best-fit dimming rates across all telescopes with multiple dates are available in the Appendix. For observations that span a significant fraction of the study period, the dimming rate can be seen to decrease in magnitude over time and fully reach zero by approximately $\sim25$ days after impact ($T+25$), contemporaneous to the return of the system to its pre-impact expected brightness. We highlight that, while there is variation from telescope to telescope, they all show a similar trend and the agreement in dimming rate across the datasets is remarkable. 

In addition, we found out that the narrower the aperture, the faster the dimming. Indeed, by comparing dimming rates in the Swope observations, the photometry measured with the 2.5”-radius aperture faded by $\sim0.023$ mags/day more ($\sim20\%$ faster in flux) than the 7.5”-radius aperture in the first seven days after impact. For a chosen aperture, the apparent fading also slowed with time. Photometry from the Danish 1.54-m dimmed at $0.149 \pm 0.002$ mags/day in the first week after impact but at  $0.108 \pm 0.006$ mags/day in the second week. For a typical telescope observing at approximately red wavelengths (e.g., with a filter similar to Sloan \textit{r}), a dimming rate of $\sim0.11-0.12$ mags/day in a 5.0”-radius aperture was seen in the first week after impact and the first month taken as a whole had a dimming rate near $\sim0.08-0.09$ mags/day on average. The aperture-dependent changes in dimming rate can be explained primarily through material leaving the aperture, but other intrinsic dimming effects -- like from bulk albedo changes in the ejecta or from phase-angle effects (see next paragraph) -- may also contribute at a lower level.

\subsection{Morphological Evolution}
We now discuss the broad-scale morphological evolution of the coma as inferred from deeper imaging with the larger aperture telescopes in our dataset to contextualize our photometric results. For larger aperture telescopes, more detailed examination of the evolution of the ejecta was possible through direct inspection and analysis of stacks of individual images. In Figure 3, we show a characteristic stacked image in the Sloan \textit{r} filter from the SOAR telescope on UTC September 29th to indicate the complex morphology of the ejecta and to introduce the terms we and others have used to describe the features seen in the ejecta. In Figure 4, we show a sequence of nine stacked images throughout the first month of the ejecta’s spread to showcase its general evolution.

\begin{figure}[ht!]
\plotone{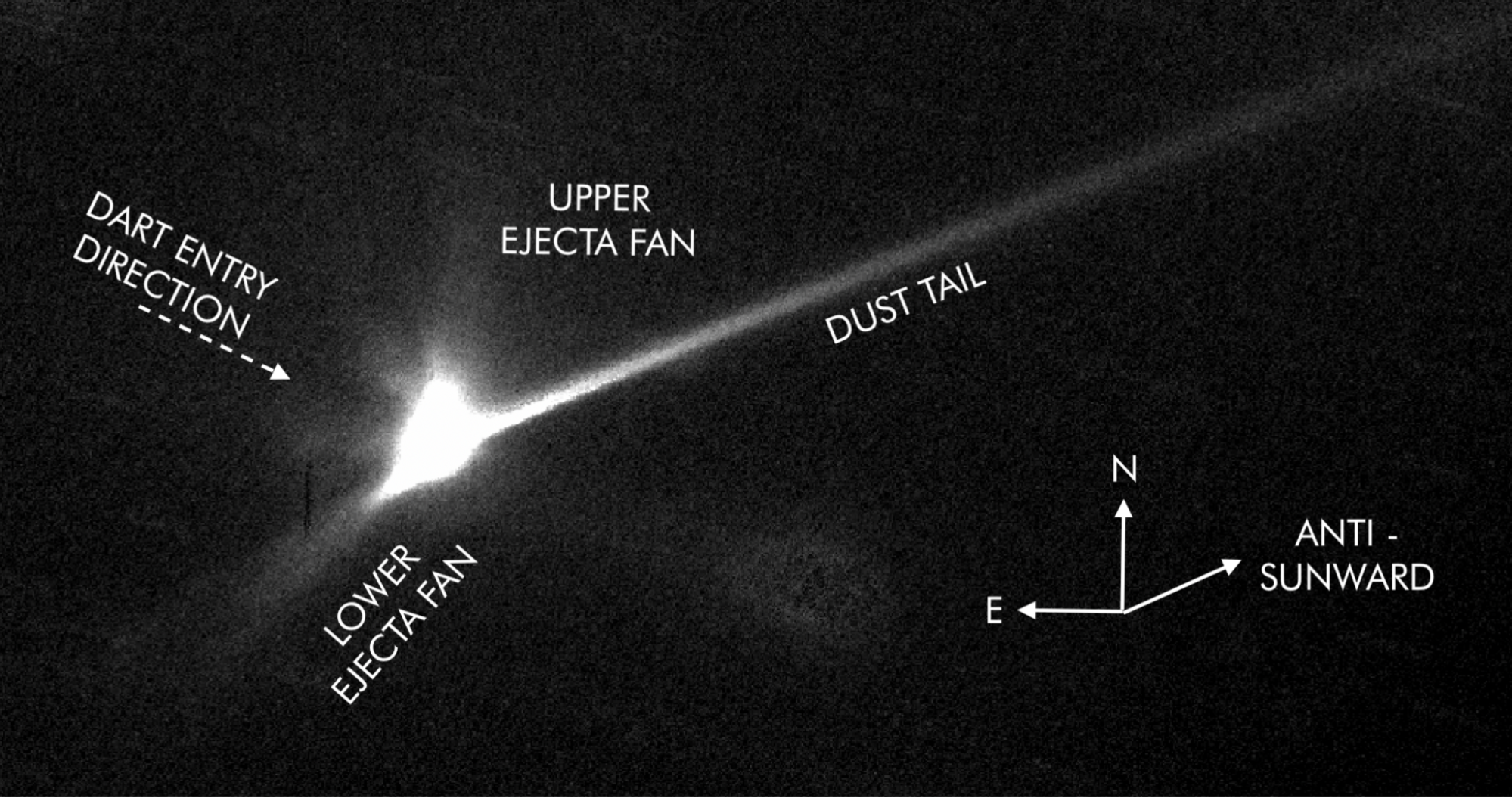}
\caption{A deep 327-second SDSS r composite image of the DART ejecta taken with the SOAR telescope on UTC September 29. Didymos and Dimorphos are at the center of the brightest area at left, the ejecta fans spread towards the left in the approximate opposite of the direction of DART’s trajectory, and the tail can be seen stretching at least 10,000 km outwards in the anti-sunward direction. The features are described in the text and shown to develop in Figure 4, where this image becomes the top-center frame. We note explicitly that the faint structure to the South-West of the asteroids is a stacking artifact.}
\label{fig:3}
\end{figure}

\begin{figure}[ht!]
\plotone{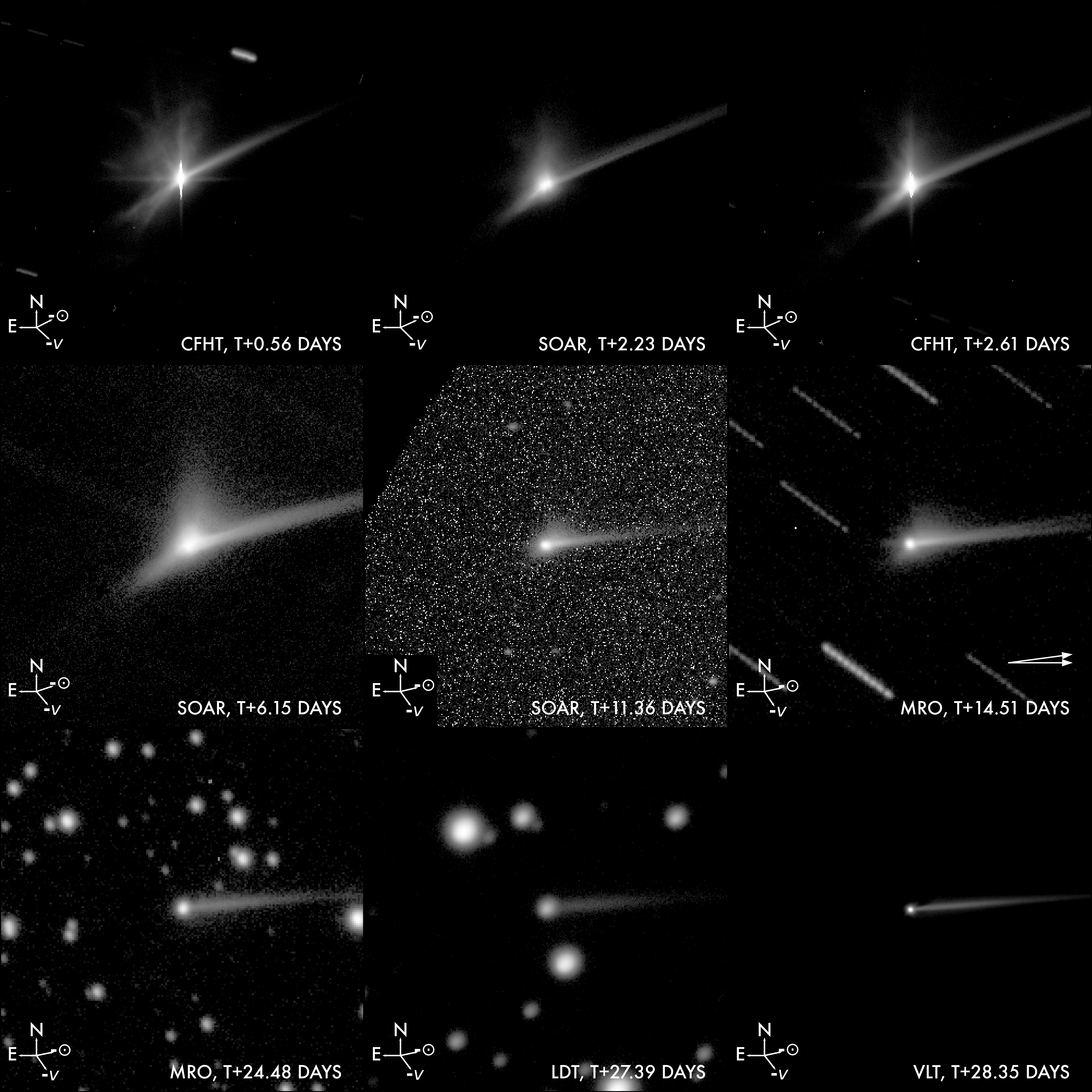}
\caption{A collection of telescopic images showcasing the evolution of the ejecta from the DART impact. The images are identically scaled and oriented (North-Up, each is 100” across) such that observations from different facilities can be compared, and a small legend in the lower left of each sub-panel shows the anti-Sunward and anti-velocity vectors. The ejecta reaches a peak in perceived complexity in the first few days, but as the ejecta fans spread out, the anti-sunward tail becomes the only clearly discernible feature by approximately day T+15 to T+20. The ‘second’ tail seen in HST observations can be discerned in data taken in good conditions (see middle panel, right, T+14.51 days, with two arrows to indicate the direction of the two tails) several days after it was first seen from space.}
\label{fig:4}
\end{figure}

Immediately after the impact, the ejecta spread out both in the direction directly opposite and slightly offset from the impact trajectory (the impact ``fans” towards the north and southeast in these images) and began to form a tail that was visible within hours (labelled as `dust tail' in Figure \ref{fig:3}). By three days post-impact, the tail was at least 10,000 kilometers long. The imaging observations were designed to maximize time spent integrating on the inner ejecta fans and coma as opposed to dithering along the tail, so in most cases the tail was detected to the edge of the detector and thus our length estimates are lower limits. Observations from the Magdalena Ridge Observatory taken on October 11th, approximately 14.5 days after impact, clearly discern the two tails first seen approximately a week earlier by HST \citep{2023Natur.616..452L} and VLT/MUSE \citep{2023A&A...671L..11O} but not seen in the October 8 SOAR data or other observations presented in this study. The detection of the transient second tail by those two studies about a week earlier than our detection of it appears to be a natural outcome of their higher angular resolution than most of our observations could provide -- as the second tail grew and became more clearly separated, it was more detectable from the ground.

\section{Discussion} \label{sec:disc}
Ground-based datasets, capable of wide wavelength coverage and long and dense time coverage, are complementary to higher resolution data sets like those taken in-situ by LICIACube \citep{2023NatCo..14.3055D} or those taken with spacebased telescopes like HST \citep{2023Natur.616..452L}. All together they form a comprehensive view of the DART impact. We first discuss the nature of how material left the Didymos system after DART’s impact, with a particular focus on the nature of the photometric change near eight days after impact, and then compare DART’s effect on Dimorphos to what was seen around other impulsive mass loss events on asteroids.

\subsection{Eight Days}
The “eight day pause” in dimming seen in at least three of our datasets occurred between visits by HST. While observations at T+5.7 days were consistent with expected dimming rates, observations at T+8.8 days were brighter than would have been expected from earlier trends \citep{2023Natur.616..452L}. Considering that we cannot detect any clear aperture-dependent timing effect in our all-night photometric averages from Swope and the Danish 1.54-m, we expect that the true onset of this brightening pause is indeed in the T+7 to T+8 day timeframe and not more than a fraction of a day earlier at most. (In Figure \ref{fig:2}, our last photometric observations prior to the `pause' are at $\sim6.3$ days after impact.) This is around the same time as the first detection of the second tail in \citet{2023Natur.616..452L} which persisted for some weeks afterwards. (\citet{2023Natur.616..452L} estimated the dust released that formed the second tail was released between 5.0 and 7.1 days after impact, so if the two events are related, the latter end of that range is preferred.) The cause of this change in photometric behavior was thus fairly rapid, only lasting approximately a day from start to finish. To investigate the nature of this brightness profile change, we convert the reduced magnitudes of Figure 1 to cross-sectional area with an assumed albedo of $p_r = 0.15$ and the phase corrections from the previously mentioned HG1G2 model \citep{2010Icar..209..542M, 2023LPICo2806.2023H} in Figure 5. We note that the explicit phase behavior of the dust, and the absolute fraction of the light that comes from the dust, is unknown and thus this approach is an approximation that could be refined in future work.

\begin{figure}[ht!]
\plotone{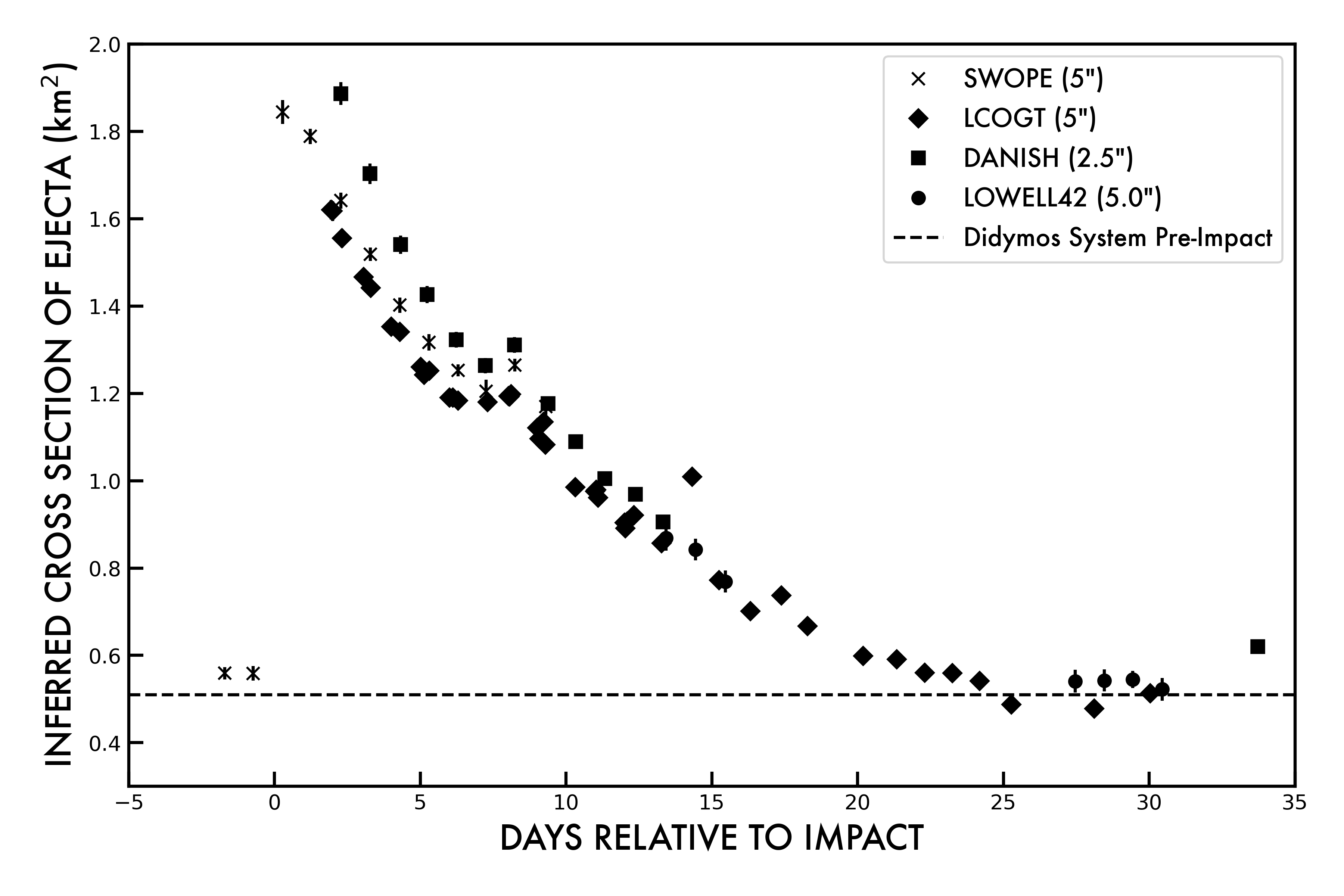}
\caption{The reduced magnitudes for a subset of the data shown in Figures \ref{fig:1} and \ref{fig:2} are converted to cross-sectional area through the same HG1G2 model as in Figure \ref{fig:1} phase function and an average albedo of $p_r = 0.15$. Accounting for the pre-impact brightness of Didymos and Dimorphos, the total cross sectional area is halved by the T+8 brightness change event.}
\label{fig:5}
\end{figure}

If the increase in cross-sectional area presented in Figure \ref{fig:5} is indeed from a secondary impact in the Didymos-Dimorphos system, we can attempt to estimate how much material was released and how this compares to the primary spacecraft impact by assuming the two events scale their cross sectional areas similarly. Accounting for the pre-pause dimming trends, the increase in cross-sectional area seen in Figure \ref{fig:5} is approximately $\sim0.1$km$^2$ to within a factor of $2\times$. By comparison, the initial increase in cross-section was considerably larger at $\sim1.3-1.4$km$^2$. If we assume that the same kinds of material were excavated from both impacts and that the bulk mass of the ejecta of an impact should scale with its cross section, we can use the estimated mass of the primary impact (a lower limit of $0.9-5.2 \times 10^7$ kg, \citealt{2023arXiv230605908R}) to estimate a mass of $0.6-4.0 \times 10^6$ kg for this proposed secondary impact. Converting to volume and considering the range of allowed densities for Didymos and Dimorphos, this extra ejecta has a volume equivalent to a sphere with a radius of $\sim3.5-8.5$m. If this is the volume of the crater excavated, then the impactor would be expected to be commensurately smaller -- and thus well within the size range of detected boulders of \citet{2023ApJ...952L..12J}, adding to the plausibility of this scenario. If a boulder with an equivalent volume disrupted -- perhaps it was weak and rapidly rotating, was impacted by another boulder, or simply fell apart on contact with the surface -- this would also add an equivalent amount of cross section to the system, but one would still need to explain the timing constraints considered later in this section, as well as justify the cross-sectional scaling argument utilized here still might apply.

If the removal of ejecta from the Didymos system was solely dominated by radiation pressure from the Sun (as opposed to dynamical effects from Dimorphos stirring up the ejecta), it would be expected that finer grains should leave the photometric aperture first. Indeed, multi-wavelength observations (see, e.g., \citealt{2023A&A...671L..11O, 2023Natur.616..461G}) show signs of the ejecta becoming redder with time, which is consistent with an increasing amount of large grains compared to smaller ones. However, there are many caveats to interpreting those trends directly, such as the length of time over which the color trends are measured or the influence of the eight day pause on bulk dimming rates. (The eight day pause could also change the interpretation of many other early measurements, e.g. whether the what was being observed was ejected in the primary impact or from a secondary process or event.) One effect of a changing grain size distribution could be a changing grain albedo distribution; larger grains are less reflective than smaller ones for Didymos-like meteorites (see, e.g., \citealt{2023PSJ.....4...52B}), suggesting that a depletion of finer grains should drive the average albedo of the ejecta lower with time, and thus provide an additional source of dimming. Moreover, the Didymos system is seen at larger phase angles later in October, meaning that a smaller fraction of each dust grain is illuminated as seen from the Earth. 

Three processes seem like plausible candidates for causing the Eight Day Pause: self-collisions among dust particles creating more material, solar radiation pressure moving existing material from ejecta fans into the dust tail, and the previously mentioned creation of new ejecta by reimpact of material inside the Didymos system. We examine each below.

If self-collisions among dust particles were important, a fragmentation cascade (larger grains breaking down into smaller ones after ejection from the impacted asteroid resulting in an increase in the total reflective cross section of the ejecta, see a discussion of this process in a different context in \citealt{2010MNRAS.409.1682T}) could explain the dimming pause or slight increase in brightness. This process would be most probable close to the asteroids where the number density of grains is highest, and thus would naturally result in a similar photometric change regardless of aperture chosen, matching our observations. However, it is hard to understand why this would happen a week after impact and not when the ejecta density is highest, i.e. minutes-to-hours after impact. Thus, while this process might have affected the photometric behavior of the system shortly after impact, it is unlikely to explain a sudden impulsive event significantly later.

A scenario discussed in \citet{2023Natur.616..452L} related to their observations of this change in photometric behavior was the radiation-pressure-driven anti-Sunward movement of material in the ejecta fans, such that some days after impact the material ejected towards the Sun would be `blown through' the photometric apertures used. This would result in an increase in the amount of dust in the aperture, and thus the amount of reflected sunlight measured, with the timing of the event being set strictly by the size distribution of material in the ejecta fans and thus the amount of time needed to change its motion sufficiently. If this were the case, the majority of the ejecta fan (one presumes the more southern ejecta fan as seen from the evolution shown in Figure \ref{fig:4}) would have had to had moved into the aperture in less than a day to be sufficiently aperture-independent to match the observations shown in Figure 2. The extent of the southern fan was several arcseconds across (corresponding to a linear distance of hundreds of kilometers, \citealt{2023Natur.616..452L}), which would require an unrealistically large speed to only affect the dimming on a single date. If the ejecta fan were moving fast and uniformly enough to enter all of the apertures on the same day, then it likely would have left them as quickly. (Changing viewing geometries might facilitate some asymmetry in onset and decay, but this is likely minor for this part of the apparition.) In that scenario, the T+9-and-later data points might be expected to lie on the pre-pause trendlines (Figure \ref{fig:2}), but the system remained systematically brighter than would be expected. These two aspects of this hypothesis are challenging to reconcile, but there is clear morphological evolution of the ejecta fans in this timeframe, so the photometric behavior in this timeframe must be modulated by this or similar processes at some level. 

A third option is the delayed re-impact of a large boulder or grouping of boulders (see, e.g., \citealt{2022PSJ.....3..118R}) ejected at T+0 during the initial spacecraft impact. Unlike the other two scenarios, this would have excavated additional material beyond that which was released in the original impact as opposed to the degradation or migration of the initial material. Material ejected directly from one of the asteroids would naturally be aperture-independent, and would dissipate and leave the system slowly over time resulting in a dimming total system brightness that is elevated over pre-pause trends. Impacts on these bodies are in general slower than the escape speed of the system, but $\sim\mu{m}$-sized particles that are lofted may be blown out of the system rapidly like much of the ejecta in the primary impact. A rapid rotation rate for whichever object was re-impacted may enhance this shedding process, given that low-speed impacts of boulders may induce surface materials to move, the smallest of which might be elevated long enough to have radiation pressure act upon them. (The movement of surface materials might very well be expected from the original impact alone -- it might take some time for the system to fully settle.) This similarity in dimming rates before and after the pause is also consistent with the injection of new material which then leaves the system similarly to the original ejecta. If the formation of the second tail and the eight day pause are related, this secondary impact scenario naturally provides an explanation for the origin and timing of the second tail. This is a critical constraint not addressed by the first two scenarios. The short duration of the dimming pause could suggest that only a narrow size range of particles was ejected due to expected low re-impact speeds \citep{2022PSJ.....3..118R}, as opposed to the wide range of particle sizes implied by the persistence of the main ejecta tail on a several-month timescale and its lack of an observed disconnection event \citep{2023PSJ.....4..138M}. The short duration of the second tail is consistent with this scenario. While a grouping of small impacts happening in short succession cannot be ruled out by our timing or aperture constraints, we are left to wonder why the impacts would be clustered in time -- a single event seems more plausible, but more modeling is needed. (Maybe the semi-stable orbits have orientations which are stable and only intersect Dimorphos's orbit at a few spots, so it's really that the boulders are clustered in space and not in time?)
In addition to a qualitative match to all of the individual attributes of our eight day pause, this secondary impact hypothesis is bolstered by the recent detection of many boulders ejected during the initial impact in deep HST imaging \citep{2023ApJ...952L..12J}. This `boulder swarm' is apparently co-moving with the asteroids, with a maximum size of $D\sim7m$ and a velocity dispersion just barely above the escape velocity of the Didymos-Dimorphos system, and would be a natural source for a secondary impactor, except that the proposed secondary impactor would have remained bound to the Didymos system rather than escaping it as the swarm did. We conclude that a secondary impact in the Didymos system is the scenario most consistent with the observations, and thus a possible source of the second tail.

The ejecta models of \cite{2018Icar..312..128Y} and \citet{2022PSJ.....3..118R} estimated that $\sim2/3$ of the re-impacts in the system hit Didymos while the rest hit Dimorphos. The distribution of velocities and times of reimpact between the two bodies is very different; most impacts on Dimorphos happen with lower velocities ($\sim0.1m/s$) and within two weeks of primary impact, while impacts onto Didymos are spread out over a longer timespan of up to months -- and thus with a wider distribution of velocities ($0.0-0.8m/s$) given that the particles have had more time to evolve onto more varied orbits. If their models are applicable to larger meter-scale boulders as seen by Hubble \citep{2023ApJ...952L..12J}, the secondary impact might have happened at low speeds onto Dimorphos, but clearly more modeling work is needed.

We note that while a secondary release of material into the system, likely driven by the impact of a boulder, is the best explanation for our data, it is not the only way to produce a transient second tail. The recent work of \citet{2023arXiv230915116K} showed that the second tail could be produced by a combination of viewing geometry and the size distribution of dust in the ejecta alone. In essence, the size-sorting nature of solar radiation pressure combined with the changing angle at which we saw the tail allowed observers to see two edges of a `cone' of material -- all released simultaneously. While it may take the re-inspection of the Didymos-Dimorphos system by the ESA Hera mission \citep{2022PSJ.....3..160M} in late 2026-early 2027 to settle the origin of the second tail for good, we note that the additional cross-sectional area added to the system $7-8$ days after impact still needs to leave somehow -- and one still needs to explain the offset in dimming produced by the eight day pause (Figure \ref{fig:2}) presumably related to the loss of that material from the system. The two origin scenarios for the second tail also do not need to be mutually exclusive, and it is not clear how the addition of a viewing geometry aspect to the second tail would change the timing relationship between the eight day pause and the second tail.

Hera's visit will thus provide an opportunity to study the frequency and importance of subsequent re-impacts, and thus a more comprehensive understanding of possible scenarios for the anomalous photometric behavior. The long-term stability of Didymos’s spin rate \citep{Nakano_sub} provides constraints on the importance of these secondary impacts prior to Hera's arrival. We note explicitly that the detection of the secondary bump in the Swope, LCOGT, and Danish datasets was only possible due to the dense temporal coverage of the system prior to the end of the first post-impact lunation, and thus other small increases in brightness or changes in photometric behavior might have occurred unnoticed. It is possible that we have detected the largest secondary impact, but several smaller impacts could have gone unnoticed behind all of the primary impact's ejecta.

\subsection{Other Activated Asteroids}
The impulsive and likely impact-driven mass loss at asteroids like (596) Scheila \citep{2011ApJ...733L...3B, 2011ApJ...741L..24I} or recurring mass loss that is not driven by sublimation like 311P/PANSTARRS \citep{2015ApJ...798..109J} or (6478) Gault \citep{2019ApJ...874L..20K, 2019ApJ...881L...6S, 2019ApJ...877L..12C}, are among the most natural comparisons to the ``activation” of Dimorphos. Scheila’s diameter is about 100 times larger than Didymos, Gault’s is 5-10 times larger, and 311P’s is slightly smaller. While 311P’s spectral type is unknown, Scheila is a carbonaceous T-type asteroid and Gault a stony S-type asteroid -- and thus plausibly a similar composition to Didymos and Dimorphos. We note that activity at the other active asteroid visited by spacecraft, (101955) Bennu (see, e.g., \citealt{2020JGRE..12506381H}) appears dominated by individual particles as opposed to tail-forming fines, and thus seems even more different still.

Scheila’s mass loss event \citep{2011ApJ...733L...3B, 2011ApJ...741L..24I} showed a complex two-fan morphology at first, which then appeared to ``drape” around that large asteroid as radiation pressure moved the finer grains anti-Sunward forming a two tail system. No additional structures or secondary brightening were noted. 311P displayed a complex multi-tailed structure that grew and persisted for months \citep{2015ApJ...798..109J}, while Gault has had episodes of mass loss and tail creation since at least 2013 \citep{2019ApJ...877L..12C}. While the processes that governed Scheila’s tail evolution appear to be the same as what has happened at Didymos, the physical scale of the separation of the two tails was much larger. The origin of the second tail -- a likely folding-over of part of the ejecta fans -- is also different than our proposed impact-driven origin for the second tail at Didymos (but is similar to the \citealt{2023Natur.616..452L} proposed origin). 

The multiple tails at 311P and the recurrent activity at Gault are plausibly driven by ongoing rotational disruption as opposed to impacts, indicating that secondary tail formation can be driven by ongoing continuous processes as opposed to strictly coming from ejecta evolution (Scheila) or other processes like secondary impacts (Didymos). Secondary impacts are implausible explanations for recurrent activity over very long periods of time, but on the hours to weeks time frame are more likely \citep{2022PSJ.....3..118R}.

We have no ability to compare the “eight day pause” in the ejecta dimming at Didymos to the other systems as no coordinated lightcurve campaign was prepared for those other targets, and thus it is plausible that post-impact photometric monitoring could have revealed more details of how those objects became and stayed active. Furthermore, with the exception of Scheila, few observations of these objects prior to discovery were accomplished. The smallest objects are often discovered active and characterized while their activity persists or recedes -- the Didymos system is the exception.

\section{Summary}\label{sec:summary}
The Double Asteroid Redirection Test (DART) spacecraft struck the asteroid (65803) Didymos's moon, Dimorphos, at 23:14 UTC on 2022 September 26th \citep{2023Natur.616..443D}, resulting in a significant amount of ejecta being released and a commensurate change in the asteroid moon's orbit. This scenario was designed \citep{2021PSJ.....2..173R} as a planetary defense exercise to understand how effective `kinetic impactor' approaches might be in future planetary defense scenarios. An important part of the overall DART mission is not just understanding how much the orbit of the moon was changed, but how and why it worked in this case and how the system evolved after impact.

In this paper, we have presented and analyzed observations of the Didymos-Dimorphos system in the first five weeks after the impact of the DART spacecraft from twelve separate observatories spanning from Antarctica to the Americas to Hawai'i. While some of the observations came from dedicated deep imaging campaigns, much of the data came from observatories involved in the lightcurve campaign to measure mutual events and thus constrain the new orbit of the moon around the primary \citep{2023Natur.616..448T}.

The system initially brightened by $\sim1.4$ magnitudes, not including the initial impact flash seen in papers like \citet{2023Natur.616..461G} and elsewhere, which then began to dim over the following weeks as material began to escape the system. Compared to a pre-impact in-situ HG1G2 model \citep{2010Icar..209..542M, 2023LPICo2806.2023H}, the system returned to its pre-impact expected brightness between $24.3$ and $25.3$ days after impact. The dimming slowed with time, with dimming rates of $0.11-0.12$ magnitudes per day in the first week after impact, slowing to $0.08-0.09$ magnitudes/day over the whole five week study period.

Eight days after impact, the dimming of the system briefly paused in three of our datasets with the highest temporal cadence (Swope, LCOGT, and the Danish 1.54-m). We constrain the timing of this photometric event to have had a rapid onset $7-8$ days after spacecraft impact, which is simultaneous or slightly after the dust that formed the second tail was released as seen in the Hubble Space Telescope dataset \citep{2023Natur.616..452L}. The system began dimming again on day $T+9$, but offset from the pre-pause trends -- the system was brighter than the pre-pause trends would have predicted. Conversion to cross-sectional area also indicates that the system's total reflective cross section increased slightly during this period. We explore several scenarios to explain this ``eight day pause," but the only scenario that can explain our photometric data is the re-impact of a small boulder released in the primary impact. The detection of many small boulders co-moving with the Didymos-Dimorphos system after impact in \citet{2023ApJ...952L..12J} adds credence to this theory. The ejecta from this secondary impact is compatible with being the origin of the second tail, but other scenarios are certainly possible \citep{2023arXiv230915116K} and more research is needed. Assuming the cross-section-to-mass ratio of the primary impact is the same as the ratio for this proposed secondary impact, we estimate a volume for this secondary ejecta as being equivalent to a sphere with a radius of $r=3.5-8.5$m, which is compatible with an origin in an impact from one of those boulders. The ESA Hera mission \citep{2022PSJ.....3..160M} will be able to search for evidence of secondary smaller impacts like that proposed here when it arrives at the system in late 2026.

The detection of the ``eight day pause" (and its likely small re-impact origin) were only possible due to Didymos's close approach to the Earth and relatively small size. That said, at a larger geocentric distance, the DART impact onto Didymos might have appeared to be easily compared to the activity onset at the similarly composed asteroid (6478) Gault, the evolution of its ejecta similar to that excavated from the asteroid (596) Scheila, and the sprouting of its second tail similar to 311P/PANSTARRS; nevertheless no single story combines them all. The impact of the DART spacecraft onto Dimorphos thus explored new ground in understanding sporadic mass loss events on the active asteroids, as well as in direct planetary defense applications. We will have to wait for the \textit{dust to settle} and for Hera to arrive at the asteroids before we understand how universal the processes that operated after the impact at Dimorphos are to other active and activated asteroids.

\begin{acknowledgments}
Much of the photometric data presented in this work is in the final stages of review for archival at the Planetary Data System's (PDS) Small Bodies Node (SBN), and a description of the lightcurve campaign these data originated in is available in \citet{Nick_subd}.

This work was supported by the DART mission, NASA Contract No. 80MSFC20D0004. The work of E.G.F. was carried out at the Jet Propulsion Laboratory, California Institute of Technology, under a contract with the National Aeronautics and Space Administration (80NM0018D0004). J.M.T.-R. acknowledges financial support from the project PID2021-128062NB-I00453 funded by MCIN/AEI/10.13039/501100011033. P.M. acknowledges funding support from CNES.
\end{acknowledgments}

\vspace{5mm}
\facilities{}

\software{}

\appendix 
\section{Observatory Information and Measured Dimming Rates} \label{sec:appendix}
In Table \ref{tab:rates}, we present the measured photometric dimming rates as measured by different facilities with different apertures over different timespans. The leftmost column lists the facility followed by the filter name followed by radius of the photometric aperture utilized. With the exception of MSU's Clear filter (closest to Johnson V in terms of response), all utilized filters have sensitivities which peak in the red. The middle column is the range of dates with respect to impact over which the rate was measured. The lowest and highest number listed for a given telescope thus indicate the range of dates that telescope acquired data for this study, e.g. 0-25 days after impact for LCOGT or 13-30 for the Lowell 42". The third table is the best-fit dimming rate in magnitudes per day as determined by a linear fits to the all-night averages described in the text with $1\sigma$ symmetric errors. The last four rows of the table, labelled `Phase Curve Only', are the dimming rates that would be expected from the HG1G2 \citep{2010Icar..209..542M, 2023LPICo2806.2023H} model described in the main text alone. While this phase curve behavior is just an approximation to the true behavior of the bulk system, it is clear that the system is always dimming significantly faster than geometry alone can account for.

\startlongtable
\begin{deluxetable*}{c|c|c}
\tablecaption{Derived Photometric Dimming Rates.}
    \label{tab:rates}
    \centering
    \startdata
        Facility (Aperture Radius) & Date Range w.r.t. Impact & Dimming Rate (mags/day)\\
        \hline
         Swope (Sloan r, 2.5") & 0-7 & $0.142\pm0.004$ \\
         Swope (Sloan r, 5.0") & 0-7 & $0.122\pm0.003$ \\
         Swope (Sloan r, 7.5") & 0-7 & $0.119\pm0.003$ \\
         LCOGT (PanSTARRS w, 5.0") & 0-7  & $0.129\pm0.002$ \\
         LCOGT (PanSTARRS w, 5.0") & 7-25 & $0.081\pm0.002$ \\
         LCOGT (PanSTARRS w, 5.0") & 0-25 & $0.086\pm0.002$ \\
         Danish 1.54-m (Johnson R, 2.5") & 0-7 & $0.149\pm0.002$ \\
         Danish 1.54-m (Johnson R, 2.5") & 7-14 & $0.108\pm0.006$ \\
         Danish 1.54-m (Johnson R, 2.5") & 0-14 & $0.115\pm0.004$ \\
         Ōtehīwai Mt. John (Johnson R, 5.0”) & 0-30 & $0.107\pm0.009$\\
         ASTEP (Johnson R, Variable) & 0-7 & $0.11\pm0.01$\\
         MSU (Clear, Variable) & 3-30 & $0.085\pm0.002$\\
         LDT (Sloan r, 5.0”) & 4-29 & $0.081\pm0.001$\\
         SOAR (Sloan r, 5.0”) & 2-25 & $0.97\pm0.003$\\
         Lowell 42" (5.0”) & 13-15 & $0.10\pm0.02$\\
         Lowell 42" (5.0”) & 13-30 & $0.040\pm0.002$\\
         Phase Curve Only & 0-10 & $0.048$\\
         Phase Curve Only & 10-20 & $0.028$\\
         Phase Curve Only & 20-30 & $\sim0.00$\\
         Phase Curve Only & 0-30 & $0.026$
    \enddata
\end{deluxetable*}


\bibliographystyle{aasjournal}
\end{CJK*}

\end{document}